# Machine Learning Model Interpretability for Precision Medicine

Gajendra J. Katuwal[*] and Robert Chen[+]

[*]Rochester Institute of Technology, Rochester, NY 14623
[+]Georgia Institute of Technology, Atlanta, GA 30332

*Abstract*— **Interpretability of machine learning models is critical for data-driven precision medicine efforts. However, highly predictive models are generally complex and are difficult to interpret. Here using Model-Agnostic Explanations algorithm, we show that complex models such as random forest can be made interpretable. Using MIMIC-II dataset, we successfully predicted ICU mortality with 80% balanced accuracy and were also were able to interpret the relative effect of the features on prediction at individual level.**

## I. INTRODUCTION

Precision medicine holds a great future in healthcare as it customizes medical care to an individual's unique disease state [1]. The widespread adoption of electronic health records has resulted in a tsunami of data that can be leveraged with machine learning based approaches in order to dissect clinical heterogeneity and aid the physician in targeted decision making.

In general, highly accurate machine learning models tend to become complex and hence are difficult to interpret. In the trade-off between predictive modelling and explanatory modelling [2], explanatory modeling is highly important among healthcare practitioners because they value the ability to understand the contribution of specific features to a model. It is very important to understand the decision process of a predictive model before its decision can be utilized in clinical setting because it affects the life and death of a patient. A predictive model has to either be interpretable, or it has to be transformed to be interpretable, in order for a user of the model to understand its decision process. Model interpretability is vital for the successful application of predictive models in healthcare, especially for data-driven precision medicine since it involves understanding a patient's unique disease state. Interpretable models would deliver actionable insights in line with precision medicine initiatives.

In this study, we perform a case study of the application of model interpretability for precision medicine on Intensive care unit (ICU) data. ICUs can benefit from the rich information that can be extracted from the improved interpretability of the models. In particular, one large area where ICU physicians and staff can benefit is in early prediction of mortality. This is a large problem because the average mortality rate at hospitals is between 8-19%, or around 500,000 deaths annually [3]. In this study, we demonstrate how a complex highly predictive model trained for ICU mortality prediction can be approximated as a simple interpretable model for each patient. Through the approximated simple models, we show that the important features' contributions during the decision process of the complex predictive model can be uniquely understood for each patietnt.

## II. METHODS

We extracted features from the Multi-Parameter Intelligent Monitoring in Intensive Care (MIMIC-II) dataset [4] containing 8,315 patients who exhibited mortality and 23,974 patients who did not. We extracted counts of medications, diagnoses, and lab tests for all patients.

We used 75% of the data for training. Remaining 25% data was used for testing. Feature selection and classification were performed using scikit-learn 0.17.1 [5]. The top predictive features were selected by ANOVA F-value feature selection test under 10-fold cross validation.

Next, a random forest (RF) [6] model with 1000 trees was trained to predict the mortality status (where value of 0 indicates no mortality and value of 1 indicates mortality). Gini impurity was used as the splitting criterion while growing decision trees. Grid tuning was performed to select the optimum number of predictors used for splitting a node of a decision tree in RF.

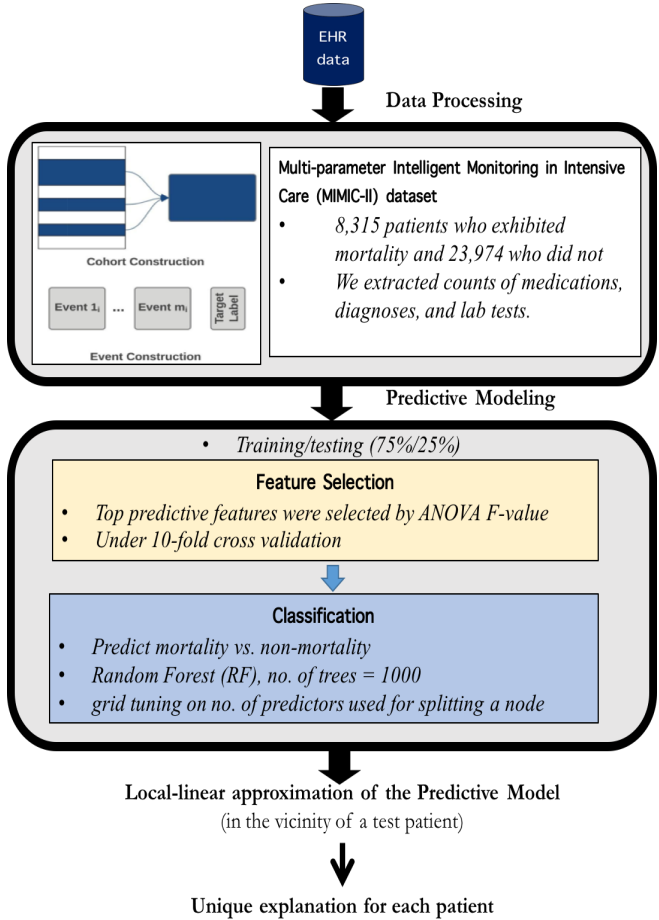

**Figure 1:** General outline of the study

Simple explanations of the contribution of important features during the decision process of the hard to interpret complex RF model were extracted using locally-interpretable model- agnostic explanations (LIME) technique [7]. The classifier decision function (the predicted probabilities by the RF model for test subjects) was subsequently fed to a LIME model. The LIME results were used to interpret the relative contribution of features for a particular patient. While RF is a complex model, LIME learns the "explanation" for an instance or a patient by approximating the RF model by a sparse liner model local to the vicinity of the patient, thus, providing a patient specific explanation.

A test subject was randomly selected to demonstrate the individual-level model interpretation derived from LIME. The non-linear decision function of the RF model is approximated by a sparse linear model in the neighborhood of the test patient (red "X"; see Fig. 2). At first, perturbed data points or instances are created around the test patient X. Then, a sparse linear model is fitted on the RF model's prediction for these perturbed instances where prediction of each perturbed instance is weighted inversely with its distance from the test patient X. Finally using the sparse linear model unique to each patient, an explanation conataining important features's contribution during the decision process of the RF model for the patient is extracted.

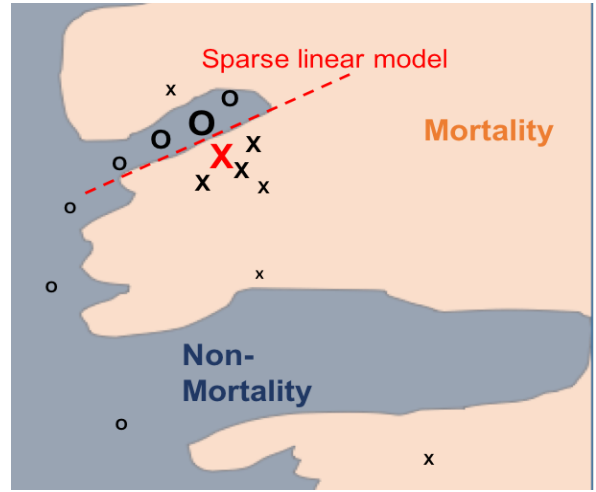

**Figure 2**: Non-linear decision function of the complex predictive model is represented by the orange/blue background. The red cross is the test patient being explained (let's call it X). Perturbed instances around X weighted by their proximity to X are fed into the model. A sparse linear model (red dashed line) is fitted for the model's prediction on these perturbed instances. This linear model approximates the non-linear decision function of the predictive model, locally in the neighborhood of X.

### III. RESULTS

Our model yielded 80% balanced accuracy in the test data. For the test subject randomly selected to demonstrate the individual-level model explanations, the top 4 most predictive features were temperature, total $CO_2$, atrial fibrillation, and lactate level (Fig. 3). Via LIME, it was identified that this particular patient with higher lactate, and more atrial fibrillation was at higher risk (78%) of mortality, which is consistent with the current medical understanding.

### IV. DISCUSSION

We accurately predicted the mortality rate of ICU

patients and were also able to uniquely identify the contribution of the important features on mortality prediction for each patient. We achieved this by combining the predictive power of RF and the patient-level model interpretability of LIME, where we linearly approximated the RF model in the patient vicinity. The explanation generated from the LIME model for our test patient is consistent and hence can be very helpful for the success of data-driven precision medicine efforts. Moreover, it also helps to answer the over-arching "*why?*" question while applying machine learning models in healthcare. The adoption of highly useful machine learning in healthcare has been delayed due to uninterpretable complex models because clinicians have not developed enough trust on these with current medical knowledge.

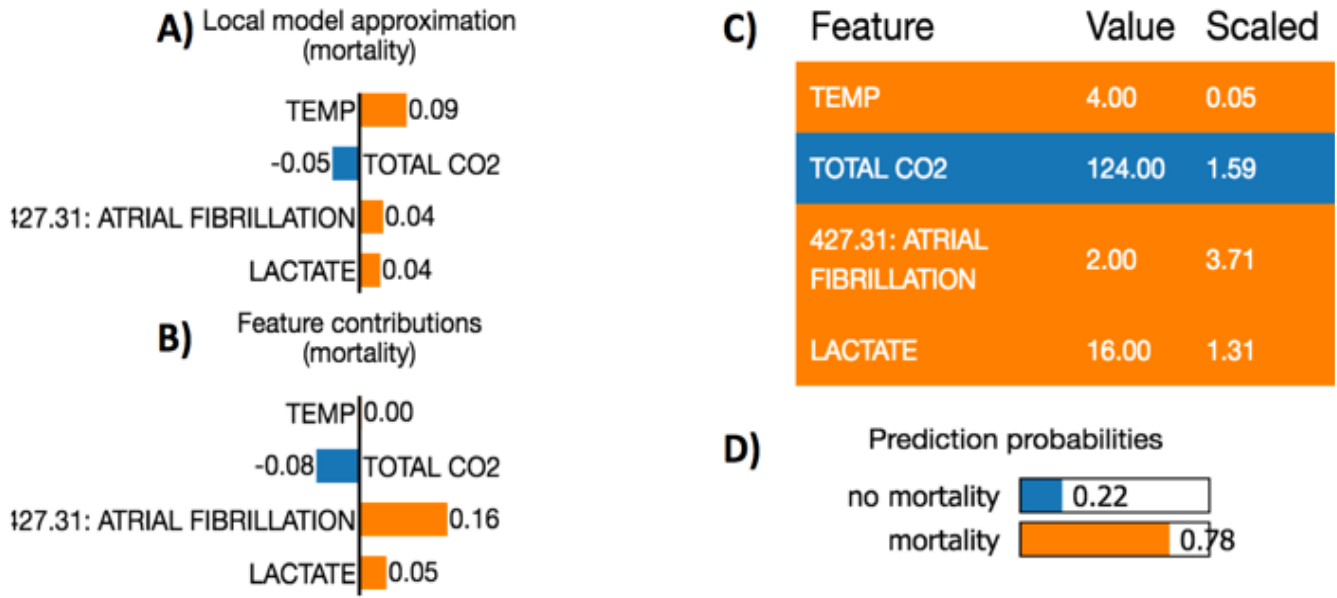

**Figure 3:** Patient specific model interpretation. A) Local model approximation in the vicinity of the patient: correlation of the features to mortality. *Temperature, atrial fibrillation, and lactate level are positively correlated with mortality.* B) Feature contributions for prediction. *Higher counts of atrial fibrillation and higher lactate level contribute towards mortality of this particular patient.* C) Value: original value for each feature and Scaled: scaled value, D) Class prediction probabilities. *The Random Forest model predicts 78% mortality for this particular test patient.*

In this case study, we successfully demonstrated that simple explanations can be extracted from a complex predictive model trained to detect occurrence of ICU mortality. As the explanations generated from the model were consistent with medical understanding, this study demonstrates that approximating complex models by simple interpretable models is one way of solving the overarching problem of uninterpretable black-box models in healthcare. In addition, approximation of the complex models uniquely for each patient provides a unique and more faithful perspective about the patient. This patient-specific model-approximation technique can be utilized to gain knowledge of each patient's unique disease state models. Using a model-approximation technique, simple and truthful explanations of the decision process of complex models can be generated. Easily understood and faithful explanations about the decision process of machine learning models can help to gain the trust of clinicians and hence accelerate the adoption of machine learning in healthcare.

## V. Conclusion

We constructed an interpretable predictive model for patient mortality using locally-interpretable model-agnostic explanation technique. We generated simple explanations from a complex model which were consistent with current medical understanding, and should motivate future work in data-driven precision medicine.